%% file: main.tex
\documentclass{IET-Conf-Paper}

\usepackage{bm}
\usepackage{graphicx}
\usepackage{subfig}
\usepackage{tikz}
\usepackage{hyperref}
\usetikzlibrary{positioning}

\def\udq{\bm{u}_\mathrm{dq}}

\def\uabc{\bm{u}_\mathrm{abc}}

\def\idq{\bm{i}_\mathrm{dq}}
\def\id{{i}_\mathrm{d}}
\def\iq{{i}_\mathrm{q}}
\def\iabc{\bm{i}_\mathrm{abc}}

\def\psidq{\bm{\psi}_\mathrm{dq}}

\def\de{\mathrm{d}}

\def\ddt{\frac{\de}{\de t}}

\begin{document}

\title{Joint identification of permanent magnet synchronous machine and inverter}

\author{Giulio Montecchio\ad{1}\corr, Sven Reimann\ad{1}, Benjamin Hartmann\ad{1}, Maximilian Manderla\ad{1}, Jan Achterhold\ad{1}, Daniel Görges\ad{2} } 

\address{\add{1}{Robert Bosch GmbH, Renningen, Germany}
\add{2}{RPTU University Kaiserslautern-Landau, Germany}
\email{giulio.montecchio@de.bosch.com}}


\begin{abstract}
In electric drive modeling, identifying the magnetic flux maps is essential for predicting accurately the torque, parameterizing a controller for tracking the torque or creating a simulation model. However, the voltage output by the controller (commanded voltage) is usually disturbed by non-linearity of the inverter, which needs to be taken into account. This paper presents a novel approach to enhance the offline flux identification from commanded voltage inputs, circumventing the need for prior identification of inverter parameters. In the dq reference frame, the flux maps are represented as static relationships between dq fluxes and dq currents. We utilize a tensor product spline model to accurately capture saturation and cross-saturation effects. The effects of the voltage disturbance on the estimated flux maps are not negligible. It is demonstrated that a joint identification of the flux maps and a simple model of the inverter can highly improve the flux model. The method is validated on FEM simulation and test-bench data.
\end{abstract}

\maketitle

\let\thefootnote\relax\footnotetext{\noindent \href{https://doi.org/10.1049/icp.2025.2025}{https://doi.org/10.1049/icp.2025.2025} \\ \textcopyright\ The Institution of Engineering \& Technology 2025}

\section{Introduction}
Accurate modeling of Permanent Magnet Synchronous Machines (PMSMs) is critical for various tasks, such as simulation or model predictive control. A key element of this modeling is the identification of the magnetic flux maps. Typically, this quantity is not directly measured, but inferred from its role in the dq frame voltage equation. However, offline identification based on collected data is susceptible to voltage disturbances caused by inverter nonlinearities. Prior work has addressed the impact of these disturbances on linear flux maps in steady-state experiments \cite{Liu2012} and proposed an experiment design for such conditions \cite{Liu2018}, or compensate it before the identification through known inverter parameters, e.g. \cite{Mink_Kubasiak_Ritter_Binder_2012}. This paper investigates the identification of a more general flux map model under transient conditions and approaches the inverter disturbance without relying on offline inverter characterization. By jointly estimating the flux maps and a simplified inverter model, the effects of the inverter can be distinguished from the intrinsic flux map characteristics, leading to a more accurate flux map model. This approach for identifying flux maps does not require inverter compensation or prior inverter parameter identification. The subsequent sections detail the mathematical model, describe the implemented identification process and illustrate the results.

\section{Modeling of the electric drive}
Flux identification in electric machines aims to model the nonlinear relationship between stator currents ($\idq$) and magnetic fluxes ($\psidq$) in the dq frame. For brevity, the subscript "dq" (e.g., $\bm{x}_\mathrm{dq}$) denotes a column vector with components along the direct ($x_\mathrm{d}$) and quadrature ($x_\mathrm{q}$) axes. We use boldface ($\boldsymbol{x}$) for vectors and capitalized boldface ($\boldsymbol{X}$) for matrices.  The linear representation of the flux $\psidq$ as the sum of a permanent magnet component $\psi_\mathrm{m}$ along the d axis and a linkage component as the product of the inductance matrix $\bm{L}_\mathrm{dq}$ and the currents $\idq$ is widely accepted and utilized. However, it does not represent saturation and cross-saturation effects present in real machines. Consequently, various alternative models have been proposed, as for example in \cite{Liu2018} or \cite{Bedetti2016}. As discussed in \cite[p. 37]{Kuhl2021}, the traditional separation of permanent magnet flux and inductance flux becomes less beneficial when considering saturation, as the permanent magnet flux can no longer be considered constant. Therefore, this paper uses the term flux map rather than inductance, recognizing that they essentially represent the same underlying quantity. We define the flux maps as general nonlinear vector functions 
\begin{equation}
\label{eq:psidq_as_f_of_variables}
\psidq = \boldsymbol{f}_\mathrm{dq}(\idq).
\end{equation}

Because $\psidq$ is typically not directly measured, identifying the flux maps relies solely on observing their effect within the dq frame voltage equation.

\subsection{PMSM Model}

The dq frame voltage equation for the PMSM is
\begin{equation}
\label{eq:continous_dq_flux_ODE}
    \Bar{\bm{u}}_\mathrm{dq} = R \idq + \omega \boldsymbol{J} \boldsymbol{f}_\mathrm{dq}(\idq) + \frac{d}{dt} \boldsymbol{f}_\mathrm{dq}(\idq),
\end{equation}
where
\begin{itemize}
    \item $R$ is the stator resistance,
    \item $\omega$ is the electrical angular frequency,
    \item $\Bar{\bm{u}}_\mathrm{dq}$ is the dq frame terminal voltage,
    \item $\boldsymbol{J} = \begin{bmatrix} 0 & -1 \\ 1 & 0 \end{bmatrix}$.
\end{itemize}
In the following, we assume a reasonably well-tuned and stable current controller is employed, enabling motor operation even with voltage disturbances. Although the electrical angle $\varphi_\text{el}$ does not explicitly appear in \eqref{eq:continous_dq_flux_ODE}, it is required for the Clarke-Park transformation, which converts commanded $\uabc$ and measured $\iabc$ quantities into the dq frame variables $\idq$ and $\udq$. For this study, a resolver is available. The resolver provides measurements of the electrical angle $\varphi_\text{el}$ and an estimate of the angular speed $\omega$. \par
Given the flux maps \eqref{eq:psidq_as_f_of_variables} that depend on $\idq$ only, their derivative in \eqref{eq:continous_dq_flux_ODE} can be computed as a total derivative. Therefore, we get the differential equation in the current
\begin{equation}
    \label{eq:continous_dq_current_ODE}
    \ddt \idq =  \boldsymbol{D}_f^{-1}(\idq) \left[ - R \idq - \omega \boldsymbol{J} \boldsymbol{f}(\idq) + \Bar{\bm{u}}_\mathrm{dq}\right] 
\end{equation}
where $\boldsymbol{D}_f$ is the Jacobian of the flux maps with respect to $\idq$, namely
\begin{equation}
\boldsymbol{D}_f(\idq) = \begin{bmatrix} \frac{\partial{f_\mathrm{d}(\idq)}}{\partial{\id} } & \frac{\partial{f_\mathrm{d}(\idq)}}{\partial{\iq} } \\ \frac{\partial{f_\mathrm{q}(\idq)}}{\partial{\id} } & \frac{\partial{f_\mathrm{q}(\idq)}}{\partial{\iq} } \end{bmatrix}.
\end{equation}
\subsection{Flux Model}
The flux is modeled using a tensor product of splines \cite[Section 8.2.1]{Fahrmeir2013}. Tensor product splines offer advantages such as regularization during learning, locality, recursive computation, straightforward analytical derivation, and linearity in the parameters. The d- and q-axis mappings can be expressed as
\begin{align}
    & f_\mathrm{d}(\idq) = \bm{b}_\mathrm{d}(\idq) \bm{\gamma}_\mathrm{d}  \\
    & f_\mathrm{q}(\idq) =  \bm{b}_\mathrm{q}(\idq) \bm{\gamma}_\mathrm{q}
\end{align}
where $\bm{\gamma}_\mathrm{d}$ and $\bm{\gamma}_\mathrm{q}$ are column arrays of parameters to be identified and $\bm{b}_\mathrm{d}(\idq)$ and $\bm{b}_\mathrm{q}(\idq)$ are the row arrays of activated basis functions, and  depend only on the currents $\idq$. The number of basis functions is an hyperparameter of the chosen flux model.
For notation brevity, we can still group the two mappings in a single dq array form as
\begin{equation}
\label{eq:splines}
    \bm{f}_\mathrm{dq}(\idq) = \boldsymbol{B}(\idq) \bm{\gamma} ,
\end{equation}
where
\begin{equation}
    \bm{\gamma} = \begin{bmatrix}
        \bm{\gamma}_{d} \\
        \bm{\gamma}_{q}
    \end{bmatrix}
\end{equation}
and
\begin{equation}
    \boldsymbol{B}(\idq) = \begin{bmatrix}
        \boldsymbol{b}_\mathrm{d}(\idq) & \boldsymbol{0}        \\
        \boldsymbol{0} & \boldsymbol{b}_\mathrm{q}(\idq)
    \end{bmatrix}.
\end{equation}

It is important to note that while the approach presented here is demonstrated using this specific model, it is not limited to it. Generalization to other linear-in-the-parameters methods, such as the one in \cite{Liu2018}, is straightforward.

\subsection{Inverter Model}
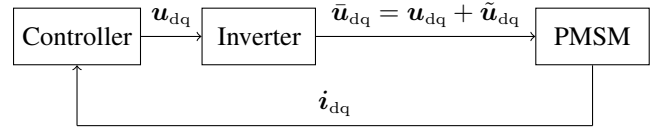
\begin{figure}[h]
    \centering
    \input{VoltageSequence}
    \caption{Voltage disturbance schematic.}
\label{fig:voltage_notation}
\end{figure}

In practice, the terminal voltage $\Bar{\bm{u}}_\mathrm{dq}$ in \eqref{eq:continous_dq_current_ODE} does not correspond to the commanded voltage 
$\udq$ due to disturbances $\Tilde{\bm{u}}_\mathrm{dq}$ introduced by the inverter, as sketched in Figure \ref{fig:voltage_notation}.
The identification and compensation of inverter-induced effects have been extensively addressed in the literature. A common approach, focusing on the impact of disturbances on the phase output voltage, is described in \cite{Choi1996}. A more detailed physical model, which accounts for parasitic capacitance, is presented in \cite{Bedetti2014}.

Following the justification in \cite{Bedetti2014}, with the assumption that the inverter legs are identical, the voltage disturbance across each phase can be parsimoniously modeled as 
\begin{equation}
\label{eq:voltage_disturbance}
     \Tilde{\bm{u}}_\mathrm{abc} \approx u_\mathrm{e} \tanh{\left(\bm{i}_\mathrm{abc}\right)}.
\end{equation}
 This qualitative model of the disturbance depends on a single parameter, the voltage error $u_\mathrm{e}$. The voltage error summarizes various effects, such as dead-time and average voltage drops across the switch and the free-wheeling diode (see \cite{Bedetti2014} for details). 
 This simple model sufficiently improves flux map fitting in uncompensated experiments, as shown later. Identification of physical inverter parameter is beyond this paper's scope. Readers interested in compensation methods are referred to the many online/offline methods, as for example to \cite{Bedetti2016} or \cite{Kim2006}.

\subsection{Joint model of PMSM and inverter}
The effects of the inverter nonlinearities on the estimation of PMSM parameters have been studied in \cite{Liu2012}, together with the effects of disturbance of the DC-link voltage. However, the study is limited to the effect of disturbances on the parameters of the linear model and on steady state experiments. In this paper, the study is extended to the identification of a general flux map as \eqref{eq:psidq_as_f_of_variables}, and to general transient experiments. \par
As reported in \cite{Choi1996}, the inverter resistances along the diode $R_\mathrm{diode}$ and the switch $R_\mathrm{sw}$ are in series with the machine resistance. Their average  can be assimilated with the machine resistance $R$ in a single resistance
\begin{equation}
\label{eq:resistance_disturbed}
\Tilde{R} = \frac{R_\mathrm{diode} + R_\mathrm{sw}}{2} +R.
\end{equation}
This means that when we are learning the flux maps from an experiment with unknown inverter, also the resistance needs to be estimated. \par

The inverter offset due to the inverter expressed in \eqref{eq:voltage_disturbance} in the dq coordinates can be written
\begin{align}
    \Tilde{\bm{u}}_\mathrm{dq} = u_\mathrm{e} \boldsymbol{K}\left( \varphi_{\mathrm{el}}\right)
 \tanh{\left(\bm{i}_\mathrm{abc}\right)}.
    \label{eq:dq_voltage_disturbance}
\end{align}
where $\boldsymbol{K} \left( \varphi_{\mathrm{el}}\right)$ is the Clarke-Park transform
\begin{equation}
\frac{2}{3}\begin{bmatrix}\cos\left(\varphi_{\mathrm{el}}\right)&\cos\left(\varphi_{\mathrm{el}}-\frac{2\pi}{3}\right)&\cos\left(\varphi_{\mathrm{el}}+\frac{2\pi}{3}\right)\\-\sin\left(\varphi_{\mathrm{el}}\right)&-\sin\left(\varphi_{\mathrm{el}}-\frac{2\pi}{3}\right)&-\sin\left(\varphi_{\mathrm{el}}+\frac{2\pi}{3}\right)\end{bmatrix}.  
\end{equation}

With \eqref{eq:resistance_disturbed} and \eqref{eq:dq_voltage_disturbance}, and splitting the terminal voltage in its commanded component and disturbance component, we can rewrite \eqref{eq:continous_dq_flux_ODE} as
\begin{equation}
\label{eq:continous_dq_flux_ODE_with_disturbance}
    \bm{u}_\mathrm{dq} = \Tilde{R} \idq + \omega \boldsymbol{J} \boldsymbol{f}_\mathrm{dq}(\idq) + \ddt \boldsymbol{f}_\mathrm{dq}(\idq) -  \Tilde{\bm{u}}_\mathrm{dq},
\end{equation}
and similarly also \eqref{eq:continous_dq_current_ODE} can be extended as
\begin{align}
    \label{eq:continous_dq_current_ODE_with_disturbance}
    \ddt \idq &=  \boldsymbol{D}_f^{-1}(\idq)  \nonumber \\
    & \cdot \left[ - \Tilde{R} \idq - \omega \boldsymbol{J} \boldsymbol{f}(\idq) + \udq + \Tilde{\bm{u}}_\mathrm{dq} \right].
\end{align}

\section{Identification approach}
The identification aims to determine the resistance and the flux maps of the PMSM in the continuous-time model using measurements from an experiment. We achieve this by discretizing either the standard voltage equation \eqref{eq:continous_dq_flux_ODE} or the extended equation \eqref{eq:continous_dq_flux_ODE_with_disturbance}, which incorporates inverter effects. Subsequently, we employ an optimization procedure to find the parameters that minimize the discrepancy between the left-hand and right-hand sides of the discretized equation at each measurement point. \par
An experiment consists of an ordered sequence of $N$ samples of the quantities
\begin{equation}
\label{eq:experiment}
\{\omega^k, \varphi^k_\text{el}, \bm{u}^k_\text{abc}, \bm{i}^k_\text{abc} \} \quad  \text{with} \quad k = 1, \ldots, N.    
\end{equation}
 They are collected with a constant sampling time $T$. \par
One possible approach to discretize  \eqref{eq:continous_dq_flux_ODE_with_disturbance} is to compute its integral at each sampling interval as
\begin{align}
\label{eq:continuous_dq_flux_ODE_integral}
    \int_{0}^{T} \bm{u}_\mathrm{dq} d\tau & = \int_{0}^{T}  \Tilde{R} \idq d\tau + \int_{0}^{T} \omega \boldsymbol{J} \boldsymbol{f}_\mathrm{dq}(\idq) d\tau \nonumber \\
    & \quad + \left[ \boldsymbol{f}_\mathrm{dq}(\idq) \right]_{0}^{T} -   \int_{0}^{T} \Tilde{\bm{u}}_\mathrm{dq} d\tau.
\end{align}
Substituting the flux model \eqref{eq:splines} and the inverter model \eqref{eq:voltage_disturbance}, considering $\omega$ to be constant during a sampling instant, we can rewrite \eqref{eq:continous_dq_flux_ODE_with_disturbance} as 
\begin{align}
\label{eq:continuous_dq_flux_ODE_integral_parameters}
    \int_{0}^{T} &\bm{u}_\mathrm{dq} d\tau  = \Tilde{R} \int_{0}^{T}  \idq d\tau + \omega \boldsymbol{J} \int_{0}^{T}  \boldsymbol{B}(\idq) d\tau  \bm{\gamma} \nonumber \\
    & + \left[ \boldsymbol{B}(\idq) \right]_{0}^{T} \bm{\gamma} -   u_\mathrm{e} \int_{0}^{T}  \boldsymbol{K}\left( \varphi_{\mathrm{el}}\right)
 \tanh{\left(\bm{i}_\mathrm{abc}\right)} d\tau.
\end{align}
 Each integral quantity is then to be either computed exactly, if possible, or approximated with the so-called quadrature rules.
 The integrals in this work are computed as 
 {\allowdisplaybreaks
\begin{alignat}{6}
    & \int_{0}^{T}\udq d\tau = \bm{u}_\mathrm{abc}^{k}\int_{0}^{T}\boldsymbol{K}\left( \varphi^{k} + \omega^k\tau\right)d\tau  \label{eq:udq_quadrature}\\
    &\int_{0}^{T}\idq d\tau  \approx  \int_{0}^{T}\boldsymbol{K}\left( \varphi^k + \omega^k\tau\right) \nonumber \\
 &\quad \quad \quad \quad \cdot \left( \iabc^{k+1} + \frac{\iabc^{k+1}-\iabc^k}{T}\tau \right) d\tau \label{eq:idq_quadrature}\\
   & \int_{0}^{T}\boldsymbol{B}(\idq) d\tau \approx \frac{T}{2} \left( \boldsymbol{B}(\idq^{k+1}) + \boldsymbol{B}(\idq^{k}) \right) \label{eq:B_quadrature}\\
 &\int_{0}^{T}  \boldsymbol{K}\left( \varphi_{\mathrm{el}}\right)
 \tanh{\left(\bm{i}_\mathrm{abc}\right)}  d\tau  \approx \nonumber \\
 & \quad\frac{T}{2} \left(  \boldsymbol{K}\left( \varphi^{k+1}_{\mathrm{el}}\right)
 \tanh{\left(\bm{i}^{k+1}_\mathrm{abc}\right)} 
 + \boldsymbol{K}\left( \varphi^{k}_{\mathrm{el}}\right)
 \tanh{\left(\bm{i}^{k}_\mathrm{abc}\right)} \right). \label{eq:disturbance_quadrature}
\end{alignat}
}
Different quadrature rules are used:
\begin{itemize}
    \item \eqref{eq:udq_quadrature}: Zero-order hold on $\uabc^k$ allows explicit integration of the Clarke-Park transformation.
    \item \eqref{eq:idq_quadrature}: A piecewise linear approximation of $\iabc$ is integrated with the Clarke-Park transformation.
    \item \eqref{eq:B_quadrature} and \eqref{eq:disturbance_quadrature}: Midpoint approximation.
\end{itemize}
 The error introduced by these approximations depends on the sampling time $T$. It is now possible to use the sum of the differences between the left hand term and right hand term in \eqref{eq:continuous_dq_flux_ODE_integral_parameters} between each couple of consecutive measured points $k$ and $k+1$ as a cost function for the identification of the unknown parameters $\Tilde{R}$, $\bm{\gamma}$ and $u_\mathrm{e}$. 
 For the sake of simplicity, we denote at each time instant $k$
 \begin{align}
& \boldsymbol{v}^k = \int_{0}^{T} \bm{u}_\mathrm{dq} d\tau \\
& \boldsymbol{q}_1^k = \int_{0}^{T} \idq d\tau \\
& \boldsymbol{Q}_2^k = \omega \boldsymbol{J} \int_{0}^{T} \boldsymbol{B}(\idq) d\tau \\
& \boldsymbol{Q}_3^k = \left[ \boldsymbol{B}(\idq) \right]_{0}^{T} \\
& \boldsymbol{q}_4^k = - \int_{0}^{T} \boldsymbol{K}\left( \varphi_{\mathrm{el}}\right) \tanh{\left(\iabc\right)} d\tau
\end{align}

Then, thanks to the linearity in the parameters of the chosen models, \eqref{eq:continuous_dq_flux_ODE_integral_parameters} can be rewritten as
\begin{equation}
\label{eq:single_row}
\boldsymbol{v}^k = \Tilde{R} \boldsymbol{q}_1^k +  \left(\boldsymbol{Q}_2^k + \boldsymbol{Q}_3^k\right) \bm{\gamma} + u_\mathrm{e} \boldsymbol{q}_4^k
\end{equation}
and the optimization problem can be expressed as a least square regression
\begin{equation}
\label{eq:ls_regression}
\begin{bmatrix}
\boldsymbol{v}^1 \\
\boldsymbol{v}^2 \\
\vdots \\
\boldsymbol{v}^{N-1}
\end{bmatrix}
=
\begin{bmatrix}
\boldsymbol{q}_1^1 & (\boldsymbol{Q}_2^1 + \boldsymbol{Q}_3^1)  &  \boldsymbol{q}_4^1 \\
\boldsymbol{q}_1^2 & (\boldsymbol{Q}_2^2 + \boldsymbol{Q}_3^2) & \boldsymbol{q}_4^2 \\
\vdots & \vdots & \vdots \\
\boldsymbol{q}_1^{N-1} & (\boldsymbol{Q}_2^{N-1} + \boldsymbol{Q}_3^{N-1}) & \boldsymbol{q}_4^{N-1}
\end{bmatrix}
\begin{bmatrix}
\Tilde{R} \\
\bm{\gamma} \\
u_\mathrm{e}
\end{bmatrix}.
\end{equation}
Since \eqref{eq:single_row} represents a two-equation system, each time step in the regression \eqref{eq:ls_regression} contributes two rows to the regressor matrix.
Notice that the joint identification of inverter and flux maps is enabled by the last column of the regressor matrix. Without it, the estimation of the resistance and the flux map parameters would suffer the omitted variable bias.\par

\section{Results}
 The validity of the approach is confirmed both with the simulation of a detailed FEM-based model including harmonic effects and with test-bench data. To assess the improvement obtained from the joint identification, Tables \ref{tb: NRMSE simulation}, \ref{tb: MAD simulation} and \ref{tb: MAD test-bench} report two rows; The first one refers to the identification of the flux parameters and resistance, the second row to the joint identification of inverter, flux parameters and resistance. There are two metrics:
\begin{itemize}
    \item Normalized Root Mean Square Error (\textbf{NRMSE}) with respect to the ground truth flux maps, and normalized with the maximum value of the fluxes. The ground truth quantities are available from the FEM model of the simulated PMSM but not on the test-bench.
    \item Mean Absolute Deviation (\textbf{MAD}) of the simulated current from the measured one.
\end{itemize}
Given an experiment in the form of \eqref{eq:experiment}, the simulated current $\Hat{\bm{i}}_\mathrm{dq}$ is obtained from integration of \eqref{eq:continous_dq_current_ODE_with_disturbance} starting form a known initial condition $\bm{i}_\mathrm{dq,0}$ and using identified parameters $\Tilde{R}$, $\bm{\gamma}$, $u_\mathrm{e}$.  At each time instant $k$, the simulation is fed with $\udq^k$ and $\omega^k$ from the experiment. The MAD can then be computed from the difference between the simulated current $\Hat{\bm{i}}_\mathrm{dq}$ and the measured $\idq$. \par 
In this study, we model the flux maps using tensor product splines. Specifically, we use 8 cubic splines, equally distributed, for both the d-axis current ($i_\mathrm{d}$) and the q-axis current ($i_\mathrm{q}$). This results in a total of 128 parameters in $\boldsymbol{\gamma}$ (64 parameters for each flux component, $\psi_\mathrm{d}$ and $\psi_\mathrm{q}$). \par
The experiments consist of a designed trajectory in the dq current disc, limited by the maximum nominal current $I_\text{max}$, which is followed at constant motor speed (that is, $ 500 \text{ rpm}$). As a trajectory, a constrained Lissajous (also known as double-sine sweep) is chosen. The current references for such trajectory are
\begin{align}
i_{\mathrm{q},\mathrm{ref}} &= I_{\mathrm{max}} \sin\left({\Phi_\mathrm{q}} t\right) \\
i_{\mathrm{d},\mathrm{ref}} &= \sqrt{I_{\mathrm{max}}^2 - i_{\mathrm{q},\mathrm{ref}}^2} \sin\left(\Phi_\mathrm{d} t\right),
\end{align}
where the angular frequencies $\Phi_\mathrm{d}$ and $\Phi_\mathrm{q}$ are design parameters of the trajectory. In this paper, the experiments for the identification data last 10 s, with a sampling time of $T = 0.1$ ms.
These experiments thoroughly explore the \textit{ dq}-current disc. As a remark, an optimal design of experiment may be crucial for the choice of electrical speed $\omega$ and the Lissajous frequencies, but this is out of the scope of this work. \par

\begin{table}[h]
\caption{NRMSE of flux maps in simulation.}
\begin{center}
\begin{tabular}{l l l}\toprule
\textbf{Approach} & \textbf{d-flux NRMSE} & \textbf{q-flux NRMSE} \\\midrule
 PSM Identification & 10.95\% & 18.19 \%\\ 
  Joint identification & \textbf{2.74}\% & \textbf{4.18}\%  \\ \botrule
  \label{tb: NRMSE simulation}
\end{tabular}
\end{center}\vspace*{-18pt}
\end{table}

\begin{table}[h]
\caption{MAD of current in simulation.}
\begin{center}
\begin{tabular}{l l l}\toprule
\textbf{Approach} & \textbf{$i_\mathrm{d}$ MAD} & \textbf{$i_\mathrm{q}$ MAD} \\\midrule
 PSM Identification & 6.74 A  & 4.98 A  \\ 
  Joint Identification & \textbf{1.35} A & \textbf{1.11} A \\ \botrule
  \label{tb: MAD simulation}
\end{tabular}
\end{center}\vspace*{-18pt}
\end{table}

\begin{table}[h]
\caption{MAD of current at test-bench.}
\begin{center}
\begin{tabular}{l l l}\toprule
\textbf{Approach} & \textbf{$i_\mathrm{d}$ MAD} & \textbf{$i_\mathrm{q}$ MAD} \\\midrule
 PSM Identification & 4.15 A  & 3.68 A  \\ 
  Joint Identification & \textbf{1.42} A & \textbf{1.50} A \\ \botrule
  \label{tb: MAD test-bench}
\end{tabular}
\end{center}\vspace*{-18pt}
\end{table}

The joint identification was applied also to data from a simulation without inverter and to data from the test-bench with active compensation. In both cases the joint identification consistently estimated the same parameters as the PMSM only identification, reporting an inverter voltage error $u_\mathrm{e}\approx 0$. 

\begin{figure}[h]
\centering
\subfloat[]{\includegraphics[width=0.9\columnwidth]{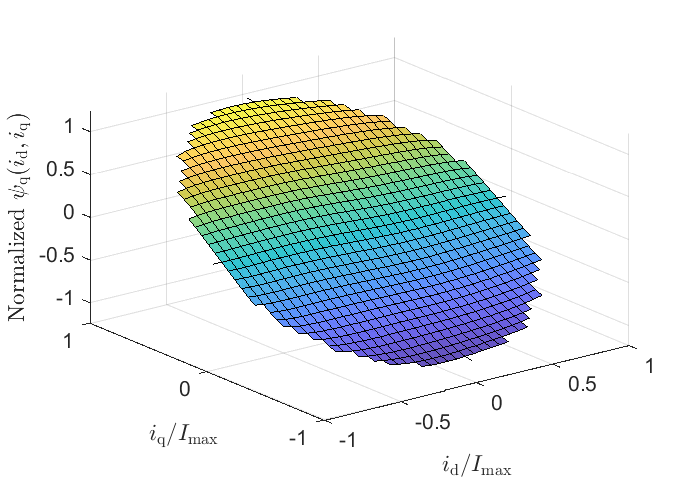} \label{fig: q_jointEstimation}}
\hfill
\subfloat[]{\includegraphics[width=0.9\columnwidth]{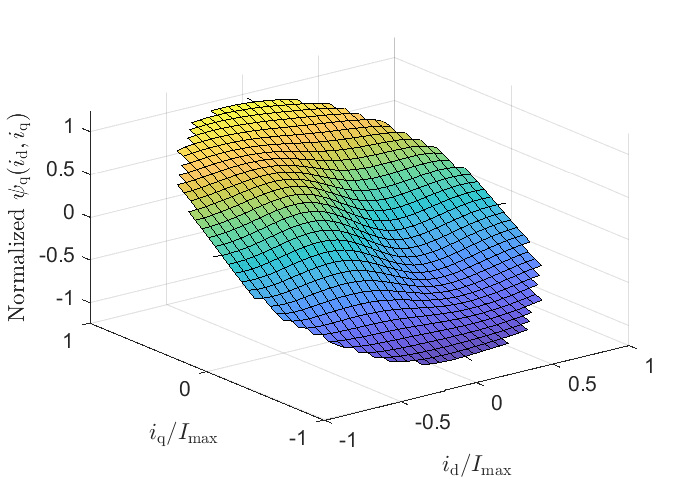} \label{fig: q_classicEstimation}}
\caption{Flux map $\psi_\mathrm{q}$ obtained with (a) joint identification and (b) without modeling the inverter effects.}
\label{fig: q_FluxComparison}
\end{figure}

\begin{figure}[h]
\centering
\subfloat[]{\includegraphics[width=0.9\columnwidth]{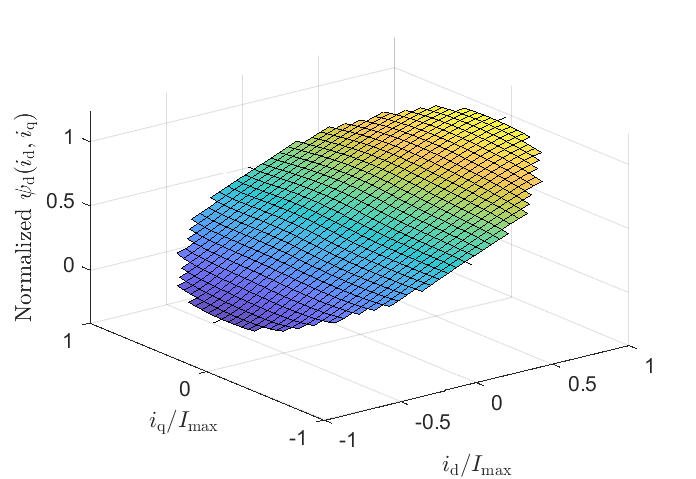} \label{fig: d_jointEstimation}}
\hfill
\subfloat[]{\includegraphics[width=0.9\columnwidth]{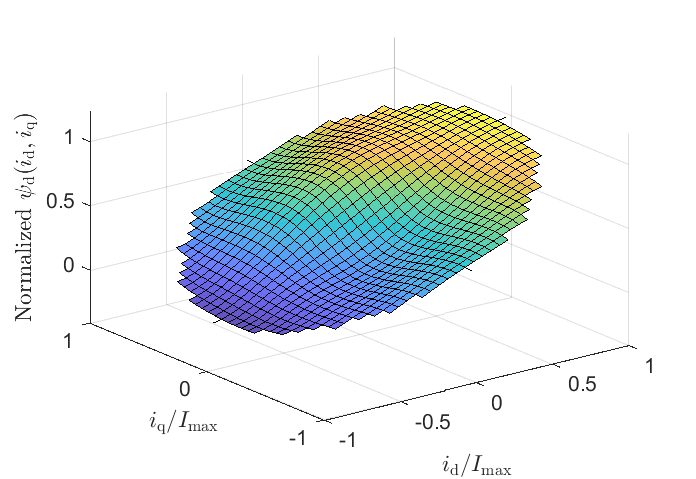} \label{fig: d_classicEstimation}}
\caption{Flux map $\psi_\mathrm{d}$ obtained with (a) joint identification and (b) without modeling the inverter effects.}
\label{fig: d_FluxComparison}
\end{figure}

The two identification approaches yield different q-axis flux maps ($\psi_\mathrm{q}$), as visualized in Figure \ref{fig: q_FluxComparison}, and different d-axis flux maps ($\psi_\mathrm{d}$), as visualized in Figure \ref{fig: d_FluxComparison}.The flux maps identified without accounting for inverter effects (Figures \ref{fig: q_classicEstimation} and \ref{fig: d_classicEstimation}) exhibits unwanted distortions, particularly at low current, due to the influence of the inverter. 

\section{Conclusion}
This paper has presented a novel approach for the offline  identification the parameters of the PMSMs from measured data and commanded voltage. By jointly estimating the flux maps, the resistance and a simplified inverter model, the proposed method effectively mitigates the impact of inverter nonlinearities on flux map accuracy, as demonstrated by the improved NRMSE and MAD values obtained in both simulations and experimental validation. Importantly, this approach does not compromise flux map identification accuracy in scenarios where voltage disturbances are already compensated. This work offers a practical and robust solution for offline PMSM modeling, eliminating the need for prior inverter parameter identification or complex compensation techniques. Future work could investigate the use of more detailed models of the inverter.

\end{document}

%% file: VoltageSequence.tex
\begin{tikzpicture}[scale=0.8] 

  \tikzstyle{block} = [rectangle, draw, text centered, minimum width=1.5cm, minimum height=0.75cm];

  \node [block] (controller) at (0,0) {Controller};
  \node [block, right=0.8cm of controller] (inverter) {Inverter};
  \node [block, right=2.9cm of inverter] (pmsm) {PMSM};
  \draw[->] (controller) -- node[above, midway, scale=1] {$\boldsymbol{u}_{\mathrm{dq}}$} (inverter);
  \draw[->] (inverter) -- node[above, midway, scale=1] {$\Bar{\boldsymbol{u}}_\mathrm{dq} = \udq + \Tilde{\boldsymbol{u}}_\mathrm{dq}$} (pmsm);

  \coordinate (x1) at ([yshift=-1cm]pmsm.south);
  \coordinate (x2) at ([yshift=-1cm]controller.south);

  \draw (pmsm.south) -- (x1);
  \draw (x2) edge[->] (controller.south);
  \draw (x1) -- (x2) node[midway, above] {$\idq$};  

\end{tikzpicture}